# A short comment on OPERA neutrino velocity measurement.

P.L. Frabetti *,  JINR, Dubna Russia
L.P. Chernenko **, Telcom MPK,  Dubna Russia

*In this report a potential problem in the data analysis of the OPERA experiment is discussed: the main issue is that the quantity "$\partial t$" used in the maximum likelihood procedure is not a "true" parameter of the parent-distribution (called PDF in the paper) but a shift in the x-axis (time scale). This means that the quantity $\partial t$ has to be considered only as systematic effect these error is not simply deducible from a gaussian distribution as stated.*

The OPERA collaboration [1] has recently reported on the early arrival time of CNGS muon neutrinos, estimating a neutrinos propagation velocity higher that the light velocity in vacuum. The later conclusion is based on some very important and not simple measurements : the distance between CERN K's and π's production target and the OPERA detector at LNGS and the clocks synchronization, again between CERN and LNGS. Both these measurements are very impressive and complicated, mainly because the general relativity has to be taken into account : see discussion in [2]. The next step in the data analysis is a statistical comparison between the time distribution of the proton pulses 1 and 2 (called PDF in the paper), extracted from the SPS at 400 Gev/c (see ref.[3] for a better understanding) and a posterori correlated to the detected neutrino events, and the time distribution of the neutrino events as measured in the OPERA detector .

\*     e-mail:  frabetti@jinr.ru
\*\*   e-mail   caiyirz@yandex.ru

In fact, this distribution is the time of the first hit in the scintillator strips related to a detected muon, occoured in a CC event into the apparatus or in the rock mountain in front of the apparatus, or related to an adronic shower, in case of NC event in the apparatus. In total ~16000 events have been observed during tree years of operation. In the OPERA paper and somewhere else [1] is explained how these distributions are obtained. We pay attention here only on the statistical aspect of the performed maximum likelihood procedure choose to estimate the δt (time difference between neutrino time and hypothetical light time in vacuum over the same distance). As far as we understood, from the [1] and from private discussion with OPERA collaborators, the PDF is the so called (in statistics language) 'parent-distribution" while the neutrino event times are considered as a sampling of the PDF distribution.

In this case, as well known from statistical books, the following function (now variables are the parameters) has to be maximized to estimate the most probable values of the parameters them-self.

$$L = \Pi_i \, PDF \, (x_i, \, \alpha, \beta, ..) \qquad (1)$$

PDF is the parent-function, $X_i$ are the measured samples and α , β etc. are parameters of the hypothesized parent-distribution; the product act over all the sample.

In the OPERA work $X_i$ is the time of the i-nth neutrino detected $T_i$ and α is δt: δt being the only parameter used in the analysis, as a term added to the independent (controlled) variable of the parent-distribution see ref.[1]

$$L = \Pi_i \, W \, (t_i + \delta t) \qquad (2)$$

First we notice that δt is not an "usual" parameter, because changing it's value does not change any of the mathematical properties (commonly called "shape") of the function L.

Then since the function has 0 (zero) value everywhere except for a small interval (~ from 0 to 11500 ns) (let us consider only the proton extraction number 1, for simplicity), where the value is almost constant (let us make such an approximation for clearness), the function is not compact. Any sampling (set of experimental values) with a single value outside the mentioned interval annul the L value (for any δt): in other word the likelihood function depend on the boundary [4]. This means there are not possible solutions (maximums) if the width of the neutrinos measured times interval (over the tree years of data collection) has some values outside the proton time interval distribution. So the convergence of the maximization procedure can occur only when the experimental values are distributed in an interval smaller than the total width of the parent distribution. This is an evident bias of the algorithm and invalidate, in our opinion, at least the claimed error of ~ 7 ns. In any case, since the proposed parameter δt looks like a systematic error of the time scale (independent variable) any attempt to consider it normally distributed has to be demonstrate . As simple solution to the issue, being the parameter δt an x-axis translation, we suggest to consider the protons and neutrinos time distributions as two "samplings" of the same parent-distribution. In this case the two estimated averages (first momentum) can be compared using the Student-distribution and consequently estimate both δt value and its error.

References:

[1]  T. Adam et al. [ OPERA Collaboration ],
"Measurement of the neutrino   velocity with the OPERA detector in the CNGS beam,"
[arXiv:1109.4897]

G. Brunetti,
Neutrino velocity measurement with the OPERA experiment in the CNGS beam,
PhD thesis, in joint supervision of the Universite' Claude Bernard Lyon-I and Universita' di Bologna, 2011,
http://operaweb.lngs.infn.it:2080/Opera/phpmyedit/theses-pub.ph

[2] Carlo R. Contaldi
The OPERA neutrino velocity results and the synchronization of clocks
[arXiv:1109.6160]

[3] J. Knoblock
Is there a neutrino speed anomaly?
[arXiv:1110.0595]

[4] Sigmund Brandt
Statistics for data analysis
North Holland Publishing Co.

P.L. Frabetti
Appunti dalle lezioni di Statistica e teoria degli errori.
Raccolta a cura degli studenti del corso di Lurea in Astronomia (UNIBO), 1985